# Domain (Grain) Boundaries and Evidence of Twin like Structures in CVD Grown Graphene

Jinho An[1,2*], Edgar Voelkl[3], Jiwon Suk[1,2], Xuesong Li[1,2], Carl W. Magnuson[1,2], Lianfeng Fu[3], Peter Tiemeijer[4], Maarten Bischoff[4], Bert Freitag[4], Elmira Popova[1], and Rodney S. Ruoff[1,2*]

[1]Department of Mechanical Engineering and the [2]Texas Materials Institute

The University of Texas at Austin, Austin, Texas

[3]FEI, Hillsboro, Oregon.

[4]FEI, Eindhoven, The Netherlands

[*]To whom correspondence should be addressed: imejin@mail.utexas.edu, r.ruoff@mail.utexas.edu

ABSTRACT

Understanding and engineering the domain boundaries in chemically vapor deposited (CVD) monolayer graphene will be critical for improving its properties. In this study, a combination of transmission electron microscopy (TEM) techniques including selected area electron diffraction (SAED), high resolution transmission electron microscopy (HRTEM), and dark field (DF) TEM was used to study the boundary orientation angle distribution and the nature of the carbon bonds at the domain boundaries. This report provides an important first step towards a fundamental understanding of these domain boundaries. The results show that, for the graphene grown in this study, the 46 measured misorientation angles are all between 11-30 degrees (with the exception of one at 7 degrees). HRTEM images show the presence of adsorbates in almost all of the boundary areas. When a boundary was imaged, defects were seen (dangling bonds) at the boundaries that likely contribute to adsorbates binding at these boundaries. DFTEM images also showed the presence of a 'twin like' boundary.

While CVD growth of graphene on Cu[1] is an excellent method for fabricating large area monolayer graphene, very little is known about the domain sizes and the domain boundary characteristics. As these boundaries likely influence the physical properties of such CVD grown graphene, understanding these boundaries is a crucial first step to ultimately engineering the boundaries to achieve desired properties for various applications. While some theoretical work has been done to understand these boundaries[2-6], very little published empirical work exists[7-9]. In this context, various TEM methods were used to better understand these domain boundaries. During the final preparation of this manuscript, an archive article[10] appeared that described the study of graphene using similar TEM techniques to that described here. However, the work described in the archive paper shows the domain (or grain, as described in the paper) sizes to be only in the order of a few hundred nanometers while the domain sizes studied in this paper are substantially larger, in excess of 10 microns, as will be shown later. This illustrates that the domain sizes can be controlled by the use of different substrates and processing conditions, which may affect the boundary characteristics and thus the overall physical properties of the graphene.

RESULTS AND DISCUSSION

The approximate domain sizes in the graphene studied here were first estimated by doing sub monolayer growth (i.e. stopping graphene growth before the Cu surface is fully covered by graphene) and observing the domain sizes in a SEM (FEI Quanta F600 ESEM), as shown in Figure 1. This showed the approximate domain sizes for the samples studied here to be on the order of slightly less than 10 microns in diameter (when the graphene fully covers the graphene, the approximate domain size can be expected to be approximately 10 microns based on the SEM image in Figure 1). By transferring (see methods) fully grown monolayer graphene samples onto silicon nitride, or Quantifoil, TEM support films that have a square array of through holes, and conducting SAED analysis, the approximate misorientation angle was identified. For a single domain graphene, one diffraction pattern will be present, namely a hexagonal spot pattern (mono and bi-layer graphene are distinguished based on the diffraction pattern intensity ratio of inner and outer spots[11]). When a SAED pattern is obtained at a domain boundary, two sets of hexagonal diffraction patterns will be present. Thus the misorientation angle here is defined as the rotation of one hexagonal set of diffraction patterns relative to the other (inset in Figure 2). Due to the six-fold symettery, a 40 degrees misorientation angle is the same as a 20 degree misorientation angle, and thus all misorientation angles are given in the range of 0 to 30 degrees. The misorientation angles were measured at up to one-tenth of a degree in accuracy. Because there is some degree of error in the misorientation angle measurements, the angles were rounded off to the nearest integer (raw data in SI). Additional error comes from the fact that even for a single domain, there are small changes in the crystal orientation due to straining or the presence of defects (see SI). The frequency distribution plot of 46 data points shows (Figure 2) that, excluding one misorientation angle of 7 degrees, all other angles range from 12 to 30 degrees. The 7 degrees misorientation angle matches a misorientation angle measured by scanning tunneling microscopy of a graphene domain boundary for graphene on copper[8]. In order to understand the significance of the distribution plot, a Kolmogorov-Smirnov[12] analysis (using

SPLUS 8.1) was performed on the data to test the hypothesis that the observed angles could be the result of a pure random selection from 0 to 30 degrees (see Figure 3 (a) in SI). In this case the hypothesis is rejected with almost 0 error, meaning that the angles cannot be the result of a random selection. In the course of obtaining the misorientation angle data, there were a few instances where a small misorientation angle was identified. In all cases, however, these were from small wrinkles or tears in the film and were thus not included in the data distribution. In order to ensure that the two sets of diffraction patterns obtained in the TEM were not due to overlapping (non A-B or A-A stacked) graphene layers, convergent beam electron diffraction (CBED) using the smallest C2 aperture was done for each case. In some cases, the CBED analysis showed that the double set of diffraction patterns were from a bilayer region (there is some adlayer present in this 'monolayer' graphene sample, albeit a small percentage), and not a domain boundary and these were also excluded from the data set. Interestingly, the approximate misorientation angle for such bilayer regions (8 regions were measured) was always close to 30 degrees; further study is indicated. The frequency distribution plot in Figure 2 also shows relatively strong peaks at 16 and 29 degrees. It is not certain at this point if these two misorientation angles are preferred due to growth mechanisms. A Kolmogorov-Smirnov analysis was done again to compare the observed angles with the uniform distribution from 12 to 30 degrees. This hypothesis was rejected with an error value of a little less than 10% (see Figure 3 (b) in SI).

For polycrystalline materials, grain boundaries are often characterized with the term "coincident site lattices" (CSL)[13]. The CSL values are important in that the values can predict the stability of the boundary and is related to the growth process including recrystallization[14]. The theoretical and experimental values of CSL values and the rotation boundaries for hexagonal systems including graphite have already been reported[15, 16]. At 14.80 degrees, an assumed CSL of 19 or 43 was reported, as well as a CSL of 13 for 29.8 degrees (the lowest CSL values from experimentally obtained rotation boundaries excluding CSL of 1 for very low angles is 13 and 19). These two angles are close to the high incidence of misorientation angles observed in this report. Other low CSL values of 1 were reported for 1.0, 1.53, and 3.0 degrees, as well as a value of 19 for 14.20 and 18.3 degrees. Theoretically, a CSL value of 7 can be achieved with a 21.8 degrees rotation and CSL of 19 for 13.2 degrees, though these were not observed experimentally. It should be noted however that in the work by Minkoff and Myron[15], only 14 experimental data points were presented. A full table of predicted and theoretical values of CSL is given in the SI. The graphene grown in this study yielded in some cases domain sizes on the order of a few tens of microns, as determined by the continuous single set of diffraction patterns obtained from the graphene over relatively long distances. To properly estimate the size of a particular domain, the domain boundary from one end to the other end needs to be identified. But because graphene transfer over large areas free of any wrinkles or tears is difficult, it was not possible to determine the approximate domain sizes. It was found in the course of the study, by doing analysis of diffraction patterns as a function of location on the graphene sample, that domain sizes in our graphene were quite large and domain boundaries were difficult to find. Generally, 10 to 20 holes in (wrinkles/tears prevent probing of a large number of adjacent holes, and thus the holes are probed where possible) the silicon nitride or Quantifoil support film (each hole with diameter on the order of a few microns) had to be probed before a domain boundary could be found. In one instance, 5 adjacent holes (thus about 20-25 microns in length from the first to the fifth) with wrinkle and tear free graphene were probed, and were found to be of a single domain. It was also typically the case that identical diffraction patterns (thus, without the occurrence of wrinkles or tears) could be obtained over 2 to 3 adjacent holes, which translates to domain sizes at least 10 to 15 microns in size. This clearly demonstrates that the domain sizes were quite large.

While the epitaxial relationship between the graphene and Cu, if present, is not yet known, the absence (with the exception of 1) of boundaries below 12 degrees suggest that if the nucleation process is purely random, neighboring islands with a low misorientation angle might undergo something similar to 'Ostwald ripening', where domain growth induces the formation of graphene with very large domain sizes. For a Cu(111) surface, it was recently reported that a weak epitaxial relationship seems to exist for graphene and the Cu,[8] but many more studies on polycrystalline Cu need to be performed. If epitaxial relationship(s) do exist for graphene and copper crystals, the non-polished multi-grain copper foil samples that we have used to date have machine marks and relatively large metal steps and thus a strongly multi-faceted surface, and different nucleation processes might be anticipated. Based on the distribution data reported here, some boundary misorientations might be favored over others, and form relatively stable boundaries.

For grain boundaries in polycrystalline materials, the boundaries are relatively flat in nature. For domain boundaries in graphene however, HRTEM images (such as the one in Figure 3) show that the boundaries are jagged. Furthermore, the boundaries seem to have a large concentration of defects. In this mask filtered TEM image, the boundary in the inset of Figure 3(a) seems to be disconnected (based on the large separation between the domains), and not forming a pentagon or heptagon type defects at this particular boundary. In other TEM images at lower magnifications, we saw that in all cases except for the area in Figure 3, the boundaries were fully covered by adsorbates (Figure 4). Perhaps there is a high concentration of defects that provide bonding sites for adsorbates to attach to at the boundaries. Such adsorbates are likely to influence the electrical and other properties of graphene; achieving even larger domain sizes in the future will allow study of the influence of boundaries and such adsorbates, if also present for those samples. Figure 3 also shows that the graphene domain on the left and the domain on the right have a different defocus. The left domain has a hexagonal crystal structure while the right does not. This is most likely due to the graphene on the right being tilted at an angle (See SI for TEM simulations) due to the strain associated with the complex bonding structure at the boundary. Imaging graphene domain boundaries in HRTEM mode provides an accurate understanding of how the carbon atoms are bonding (and not bonding) at the boundaries. While using state-of-the art microscopy facilities has provided unprecedented understanding of the crystal structure of graphene, there is difficulty in observing the domain boundaries at lower magnification. In this context, a "dirty" dark field TEM imaging technique[17] was used to observe the boundary at a lower magnification (Figure 5). By acquiring the image with an objective aperture placed over one of the diffraction spots, the two different domains were clearly identifiable and the precise location of the domain boundary observable. Surprisingly, the boundaries in another location turned out to be much more complex, where twin-like boundary structures (Figure 5) were present much like in an interdigital transducer (IDT). These types of boundaries are most likely due to the comb-like preferential growth of graphene along metal steps (SI). Because the Cu foil is a polycrystalline substrate, some of the grains will be heavily stepped depending on the crystal orientation of the surface. This suggests that controlling the Cu surface to reduce the metal steps may be beneficial in removing these types of graphene structures.

CONCLUSION

By using various TEM techniques, the approximate domain sizes and misorientation angle distributions were studied. While different growth conditions are likely to result in different boundary characteristics, the graphene studied here had domain sizes on the order of approximately 10 microns, with the largest domain found to be about 20 microns in diameter. 45 of the 46 domain boundary angles measured were in the range of 12 to 30 degrees, with one being at 7 degrees. HRTEM imaging showed a large presence of defects that lead to heavy contamination of adsorbates at these boundaries. By using DFTEM, some boundaries were observed to be 'comb-like', resulting in alternating 'A-B-A-B' twin like structures in the graphene (not to be confused with the 'AB' Bernal type stacking of layers in graphite). This shows how the graphene grown along metal step edges creates a monolayer graphene in that region. Further study of domain boundaries in samples grown in a variety of ways is indicated, and mapping of physical properties as of function of such detailed understanding of structure is called for.

METHODS

CVD graphene grown on a copper foil[1] was drop-coated with PMMA (dissolved in chlorobenzene) and dried. The sample was then floated onto an iron nitrate solution (~5 wt. %) and exposed to that solution overnight to dissolve the Cu. The PMMA/graphene film floating on the iron nitrate solution was rinsed with ultrapure water (transferred to a beaker with ultrapure water using a spatula) and transferred onto a $SiO_2$ (275nm)/Si substrate and rinsed with ultrapure water and then air dried. Additional PMMA solution was dropped onto the PMMA/graphene film to dissolve the solid PMMA and relax the graphene film[18]. After drying the additional PMMA, the substrate was placed in acetone to remove the PMMA and subsequently rinsed with water. Cu Quantifoil TEM grids were placed onto the graphene film and isopropyl alcohol dropped on top of the grids to allow the carbon film in the Quantifoil to come in contact with the graphene. The substrates were dipped into a KOH solution (~30wt. %) to slightly etch $SiO_2$ until the grids floated free off of the substrate. The grids were finally rinsed with ultrapure water and dried.

The HRTEM imaging was performed on a Titan TEM equipped with a CEOS GmbH image Cs-corrector and a Wien Filter monochromator at the FEI NanoPort facilities in Eindhoven, Netherlands, and on a Titan TEM equipped with a CEOS GmbH image Cs-corrector at the NanoPort facilities in Hillsboro, Oregon, USA, both at 80keV.

DFTEM imaging was performed at 200keV on a JEOL2010F TEM. No apparent damage to the graphene was observed at the magnification used. Objective aperture was placed over one of the diffracted beams, and the DF image acquired using an acquisition time of 300s.

ACKNOWLDGEMENTS

The authors gratefully acknowledge the support of FEI Co. for providing TEM access. The authors also gratefully acknowledge the financial support of this work by NSF Grant CMMI-0700107, the Nanoelectronic Research Initiative (NRI-SWAN; #2006-NE-1464), and the DARPA CERA grant.

Figure. 1 SEM image of sub-monolayer graphene islands on copper. White circles show areas where domains are merging.

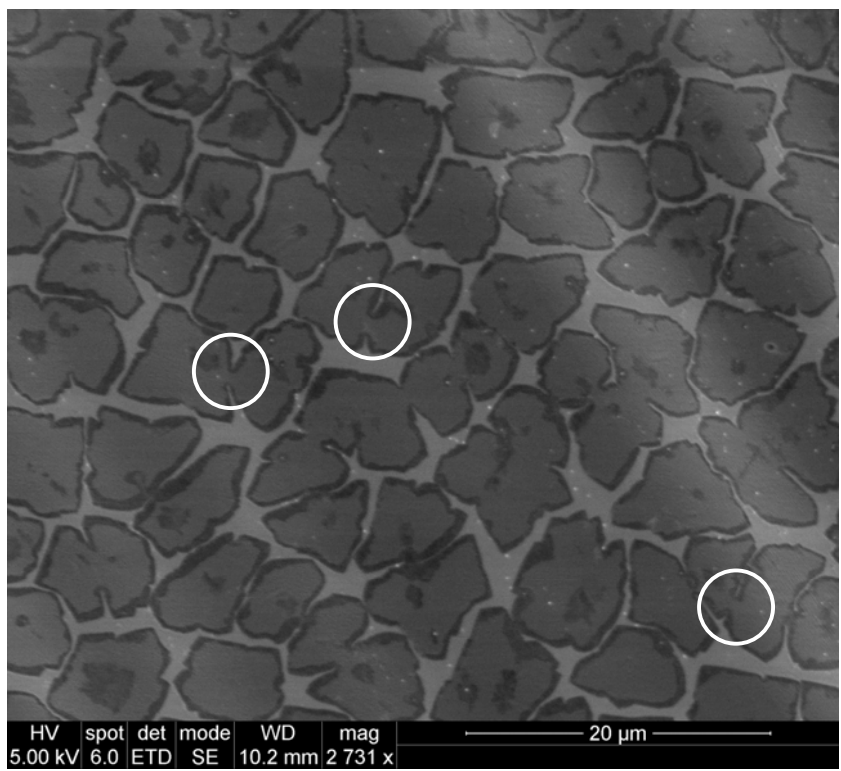

Figure. 2 Misorientation angle distribution plot of 46 data points. Diffraction pattern inset in graph shows how the misorientation angle was defined.

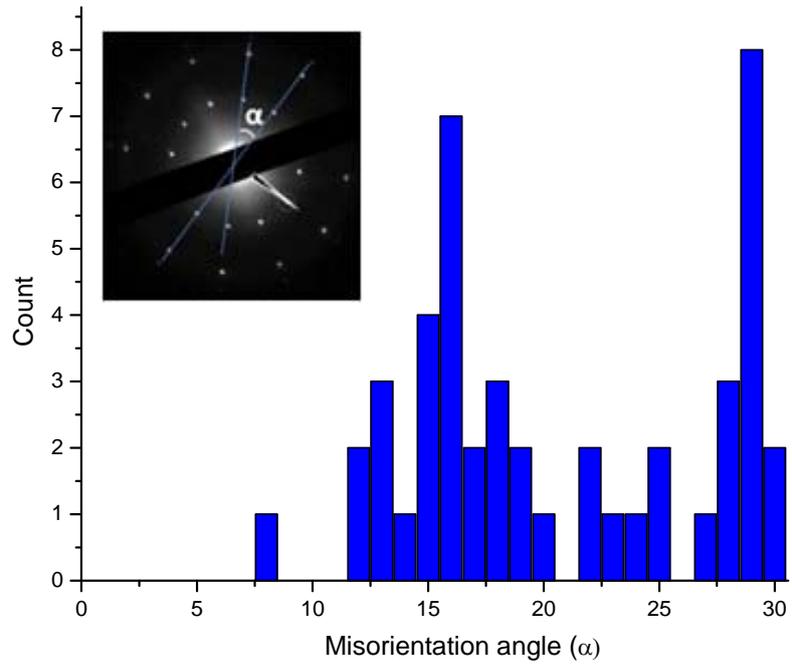

Figure. 3 (a) Mask filtered HRTEM image of graphene at a domain boundary. Enlarged image of inset in (a) shown in (b). Boundary seems to be disconnected at the upper domain's zigzag array.

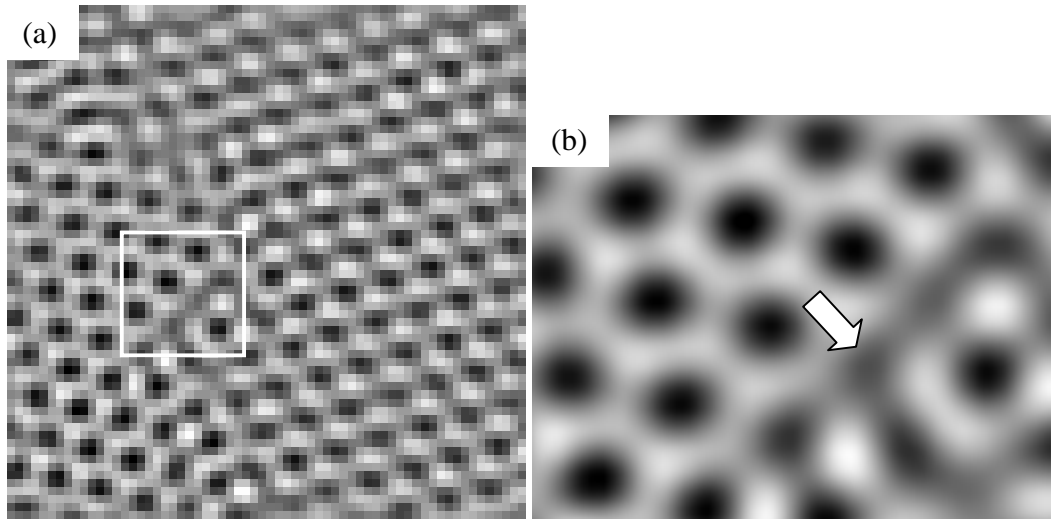

Figure. 4 TEM image of graphene at domain boundary. Blue line denotes approximate line at which the domain boundaries exist. (b) and (c) show the FFT from the left and the right area of the blue line, respectively, and the FFT in (d) is from the whole TEM image.

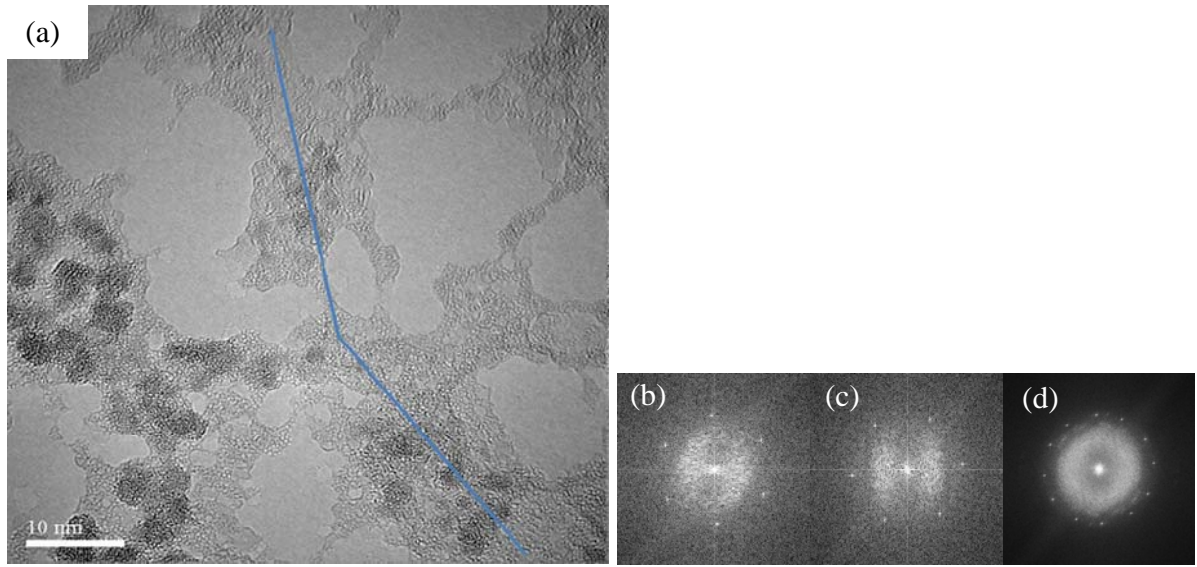

Figure. 5 (a) Bright field TEM (BFTEM) and (b) Dark field TEM (DFTEM) of the identical spot on a monolayer graphene. White arrow in (a) and (b) point to the same particle contamination on the graphene surface. (c) Two sets of diffraction pattern roated approxiamtely 25 degrees from one another. Objective aperture was placed over spot circled in (c) for DFTEM image acquisition. (d) Comb-like graphene structure shown, identified by DFTEM. Circle in (e) shows the diffraction spot used to image the graphene in (d).

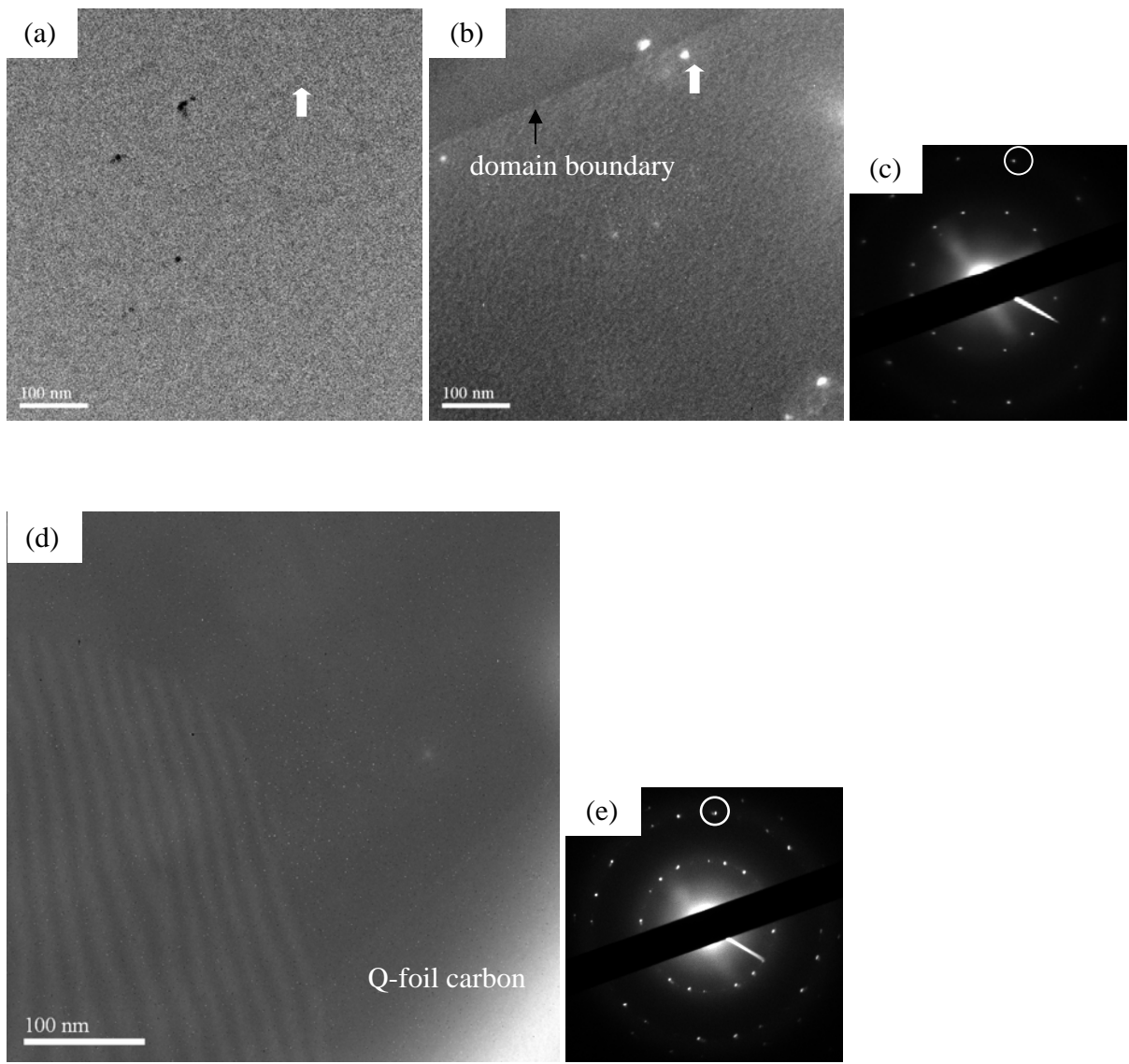